\documentclass[twocolumn,showpacs,preprintnumbers,amsmath,amssymb,APSl,prd,nofootinbib,superscriptaddress]{revtex4-1}

\usepackage{dcolumn}% Align table columns on decimal point
\usepackage{bm}% bold math
\usepackage{ifpdf}
\usepackage{hyperref}
\usepackage{dcolumn}
\usepackage{bm}
\usepackage[spanish,english]{babel}
\usepackage{amsfonts}
\usepackage{amssymb}
\usepackage{graphicx}
\usepackage[latin1]{inputenc}

\newcommand \be{\begin{equation}}
\newcommand \en{\end{equation}}
\newcommand \bea{\begin{eqnarray}}
\newcommand \ena{\end{eqnarray}}

\newcommand \Tr{\mbox{\rm Tr}}

\begin{document}

\title{Classical resolution of black hole singularities in arbitrary dimension}

\author{D. Bazeia} \email{bazeia@fisica.ufpb.br}
\affiliation{Departamento de F\'isica, Universidade Federal da
Para\'\i ba, 58051-900 Jo\~ao Pessoa, Para\'\i ba, Brazil}
\author{L. Losano} \email{losano@fisica.ufpb.br}
\affiliation{Departamento de F\'isica, Universidade Federal da
Para\'\i ba, 58051-900 Jo\~ao Pessoa, Para\'\i ba, Brazil}
\author{Gonzalo J. Olmo} \email{gonzalo.olmo@csic.es}
\affiliation{Departamento de F\'isica, Universidade Federal da
Para\'\i ba, 58051-900 Jo\~ao Pessoa, Para\'\i ba, Brazil}
\affiliation{Departamento de F\'{i}sica Te\'{o}rica and IFIC, Centro Mixto Universidad de
Valencia - CSIC. Universidad de Valencia, Burjassot-46100, Valencia, Spain}
\author{D. Rubiera-Garcia} \email{drubiera@fudan.edu.cn}
\affiliation{Center for Field Theory and Particle Physics and Department of Physics, Fudan University, 220 Handan Road, 200433 Shanghai, China}
\author{A. Sanchez-Puente} \email{asanchez@ific.uv.es}
\affiliation{Departamento de F\'{i}sica Te\'{o}rica and IFIC, Centro Mixto Universidad de
Valencia - CSIC. Universidad de Valencia, Burjassot-46100, Valencia, Spain}


\date{\today}

\begin{abstract}
A metric-affine  approach is employed to study higher-dimensional modified gravity theories involving different powers and contractions of the Ricci tensor. It is shown that the field equations are \emph{always} second-order, as opposed to the standard metric approach, where this is only achieved for Lagrangians of the   Lovelock type. We point out that this property might have relevant implications for the AdS/CFT correspondence in black hole scenarios.  We illustrate these aspects by considering the case of Born-Infeld gravity in $d$ dimensions, where we work out exact solutions for electrovacuum configurations. Our results put forward that black hole singularities in arbitrary dimensions can be cured in a purely classical geometric scenario governed by second-order field equations.
\end{abstract}

\maketitle

\section{Introduction}

A major reason to consider the physics of modified gravities in extra dimensions is provided by the Anti-de-Sitter/Conformal Field Theory (AdS/CFT) correspondence. This is a duality relating the physics (more precisely, the thermodynamics) of black holes defined in AdS space ($n$-dimensional) and the CFT lying on its ($(n-1)$-dimensional) boundary \cite{AdS/CFT}. It constitutes an example of the more general concept of gauge/gravity dualities and holography, by which the physics of the interior of a system can be described through the physics of its boundary. Even though the AdS/CFT correspondence was originally developed within string theory, it has far-reaching consequences and applications in other areas, such as in the description of strongly-interacting and condensed matter systems \cite{Condensed}. One can thus forget about the original motivation for the correspondence, focus on the analysis of properties of black holes in AdS space-times, and try to work out the corresponding aspects of the CFT using the dictionaries established in the literature \cite{Dictionaries}.

On the CFT side, one of the most interesting cases is the four-dimensional one, implying that one is led to consider gravity theories in five dimensions. An important lesson of the existing AdS/CFT dictionaries, is that new couplings and interactions in the gravitational sector added to the Einstein-Hilbert (EH) Lagrangian of General Relativity (GR) have a direct correspondence on the CFT side. By adding new elements in the gravitational sector one could thus explore new couplings in the CFTs. However, the addition of higher-order curvature invariants in the gravitational sector usually gives rise to higher-order derivative field equations. This largely prevents finding analytical solutions and usually results in the appearance of ghosts and instabilities, which casts severe doubts on the physical consistency of the corresponding theory.

A well known way out of this problem is to consider a particular combination of quadratic corrections to the EH Lagrangian that removes the undesired higher-order derivative terms, known as Gauss-Bonnet theory. In this theory, in addition, exact solutions can be obtained \cite{GB}. This theory contains a single independent parameter, which limits the range of dual CFTs that can be studied. A natural next step is to add cubic curvature terms to the Gauss-Bonnet theory in appropriate combinations so as to keep second-order field equations. This procedure can be generalized to arbitrary orders by means of the so-called Lovelock theories \cite{Lovelock}. A drawback of using the Lovelock approach to add cubic couplings is that in five and six space-time dimensions the corresponding cubic terms become topological invariants. This means that they add no new contribution to the field equations (the same applies for the Gauss-Bonnet combination in four space-time dimensions). One is thus faced with a serious limitation to include new curvature couplings in the gravitational sector which can contribute in a nontrivial manner to the field equations, which would have an impact on the dual CFT.

To circumvent this argument, a new procedure has been developed based on the addition of a set of new curvature-cubed terms to the action in five dimensions such that, though the field equations are, in general, of third-order, the linearized equations describing gravitons propagating in the AdS background are second-order \cite{quasi-topological}. They are called ``quasi-topological gravities", as the extra terms are not topological invariants any longer. However, besides the lack of naturalness in this procedure, as a matter of fact one is forced to solve highly complicated equations where solutions are only obtained under restrictive conditions. Extra freedom in these theories is thus limited and a aesthetically more appealing approach would be desirable.

In this work we consider a different approach to this problem and face it from the perspective of metric-affine (or Palatini) geometry. In the Lovelock and quasi-topological approaches, one implicitly assumes the hypothesis that the connection is metric-compatible, i.e., that it is defined \emph{a priori} in terms of the Christoffel symbols of the metric. This assumption, which is more a historical convention than an experimentally supported fact\footnote{Indeed, analysis of the physics of crystalline structures with defects on their microstructure, supports the view that non-Riemannian geometries are favoured in Nature \cite{crystalline}, which might have important consequences for the understanding on the microscopic structure of space-time \cite{orl14}.}, is at the root of the limitations of the previous approaches.  We will see that relaxing the compatibility condition between metric and connection and allowing the connection to be determined by the field equations generically produces ghost-free, second-order metric field equations. This strategy, therefore, provides a fair amount of freedom in the gravitational couplings without the restrictions of Lovelock theories and quasi-topological models.

It is worth stressing at this point that metric and connection carry very different physical meanings. While the former deals with properties locally defined in space, such as measurements of lengths, areas, volumes, etc, the latter is related to properties remaining invariant under affine transformations, such as parallelism (see \cite{Zanelli} for a pedagogical discussion). Forcing these two structures to be related to each other a priori is an unnecessary constraint that limits the potential of these theories.  In the metric-affine viewpoint, the field equations are obtained by independent variation with respect to metric and connection. As we will see, in many cases of interest, the connection turns out to be a constrained (non-dynamical) object which, in general, differs from the Levi-Civita connection (the exception to this rule corresponds precisely to Lovelock theories \cite{Exirifard,Borunda}). As GR is a particular case of Lovelock gravities, one finds that the  Palatini approach gives in this case the same results as the standard metric approach. However, in general, the mathematical and physical properties of modified gravities in the metric and Palatini approach largely differ from each other (see \cite{Olmo:2011uz} for a review on the topic).

As already mentioned, a key aspect of the Palatini formulation is that, for a large family of functional forms of the Lagrangian density, the field equations turn out to be second-order. This is so due to the active role played by the matter in the construction of the connection. In particular, in absence of matter fields, the connection becomes metric compatible and the field equations boil down to those of GR with a cosmological constant term, which avoids the presence of extra propagating degrees of freedom. When matter is present, the connection equation can be seen as defining the Levi-Civita connection of a new metric $h_{\mu\nu}$, which is algebraically related to the metric $g_{\mu\nu}$ originally appearing in the action. This algebraic relation depends on the matter sources and on the particular gravity Lagrangian chosen. Additionally, one finds that the field equations for $h_{\mu\nu}$ can be cast in Einstein-like form, which greatly simplifies their resolution.

Recently,  we have successfully applied this formalism to the case of higher-dimensional $f(R)$ Palatini theories and worked out exact solutions in the five-dimensional case \cite{blor}. Now we extend the scope of the methods developed there by considering the case of theories including also powers of the Ricci-squared scalar, $Q=R_{\mu\nu}R^{\mu\nu}$, and other curvature contractions. In particular, we consider a gravitational Lagrangian of the Born-Infeld type, which formally contains up to four powers of the Ricci tensor. We derive the field equations for these theories and write them in terms of the $h_{\mu\nu}$ metric to highlight their second-order character and how a cosmological constant naturally emerges. Focusing on Born-Infeld gravity, we work out exact solutions corresponding to an electrostatic spherically symmetric electromagnetic field. A discussion on the internal structure of its solutions and its implications is provided. In particular, we show that these solutions are self-gravitating structures with nontrivial (wormhole) topology and, at the same time, are geodesically complete and non-singular. This is so despite the presence of curvature divergences, which demands for a reconsideration of the role traditionally attributed to curvature invariants to characterize space-time singularities, since their presence in our scenario pose no obstacle for the extendibility of paths through the wormhole.

Regarding this last point, we note that one's intuition tends to establish a correlation between space-time singularities and curvature divergences. In fact, this question has been discussed in the literature from different perspectives, including the viewpoint from the philosophy of science (see \cite{Curiel2009} and references therein). However, from a mathematical perspective, the characterization of singular space-times is more subtle \cite{Geroch:1968ut,Hawking:1973uf,Wald:1984rg}. In this sense, even Minkowski space-time with a point (or small domain) removed can be regarded as singular despite the complete absence of curvature divergences. The key point is that the removal of a point (domain) has a deep impact on the existence of physical observers. If observers are regarded as point-like entities that follow geodesics, the removal of a point (domain) implies that observers reaching to that point (domain) may simply cease to exist there (or come into existence from nowhere). Since that point (domain) does not belong to the space, physical observers are simply not defined there. There is no Physics at that point (or domain). Whether an extended object can experience arbitrarily high accelerations or deformations due to curvature divergences is of secondary importance as long as it can exist. The key point, therefore, is the existence of physical observers, which are characterized by complete geodesics.

In a series of previous papers, some of us addressed the issues of geodesic completeness in black hole space-times with wormhole structure \cite{GCiWHST}, the impact of curvature divergences on extended observers (defined by geodesic congruences) \cite{Congruences}, and also the characterization of curvature divergences by means of the scattering of (scalar) waves off the effective potentials that they generate \cite{Olmo:2015dba}.  It was found that these space-times are geodesically complete, that the volume of extended objects remains finite as the divergent region is crossed, and that wave propagation is uniquely determined across the curvature divergence, being possible the computation of the reflection and transmission coefficients that characterize the barrier associated to the curvature divergence.  In this work we will extend some of those results derived in four space-time dimensions  to the higher dimensional case.

\section{Palatini theories with Ricci scalar and Ricci-squared invariants} \label{sec:II}

In order to illustrate the basic strategy to deal with Palatini theories, consider $d$-dimensional theories defined by the following action
\be \label{eq:action}
S=\frac{1}{2\kappa^2} \int d^d x \sqrt{-g} f(R,Q) + S_m(g_{\mu\nu},\psi_m)
\en
where $\kappa^2$ is a constant related to $d$-dimensional Newton's constant in appropriate system of units, $g$ is the determinant of the space-time metric $g_{\mu\nu}$, the curvature invariants $R=g_{\mu\nu}R^{\mu\nu}$ and $Q=R_{\mu\nu}R^{\mu\nu}$ are constructed with the Ricci tensor, $R_{\mu\nu}={R^\rho}_{\mu\rho\nu}$, which follows from the Riemann tensor

\be \label{eq:Riemann}
{R^\alpha}_{\beta\mu\nu}=\partial_{\mu}
\Gamma^{\alpha}_{\nu\beta}-\partial_{\nu}
\Gamma^{\alpha}_{\mu\beta}+\Gamma^{\alpha}_{\mu\lambda}\Gamma^{\lambda}_{\nu\beta}-\Gamma^{\alpha}_{\nu\lambda}\Gamma^{\lambda}_{\mu\beta}
\en
In these expressions the connection $\Gamma \equiv \Gamma_{\mu\nu}^{\alpha}$ is {\it a priori} independent of the metric $g_{\mu\nu}$. For simplicity, we assume that the connection is not coupled to the matter action, $S_m$, which depends only on the metric and on the matter fields, denoted collectively as $\psi_m$. In general, the antisymmetric parts of both the connection (called torsion), $\Gamma_{[\mu\nu]}^{\alpha}$, and the Ricci tensor, $R_{[\mu\nu]}$, can be non-vanishing. In this work, however, both contributions are neglected for simplicity (see \cite{Olmo:2013lta} for a discussion of the general case).
Under these conditions,  variation of the action (\ref{eq:action}) with respect to metric and connection yields
\bea
\delta S\!&=&\!\! \frac{1}{2\kappa^2}\!\! \int \!d^d x \sqrt{-g} \Big[\!\Big(f_R R_{\mu\nu} \!+\! 2f_Q g^{\alpha\beta}R_{\mu\alpha}R_{\beta\nu}\!\! -\frac{1}{2}g_{\mu\nu}f \Big) \delta g^{\mu\nu} \nonumber \\
&&+ \Big(f_R g^{\mu\nu}+2f_Q R_{\alpha\beta}g^{\alpha\mu}g^{\beta\nu} \Big) \delta R_{\mu\nu}(\Gamma) \Big]+ \delta S_m
\ena
where we have used the short-hand notation $f_X \equiv \partial f/\partial X$. An important point now is to realize that the  Ricci tensor is an object  constructed only with the connection and, therefore,  \emph{does not depend on the metric} $g_{\mu\nu}$. This point is crucial to understand the different structure of the field equations as compared to the (more standard) metric approach. In the Palatini formulation, we have to write the variation of $R_{\mu\nu}$ in terms of the variation of the connection. To do this we make use of the relation\footnote{For simplicity we omit the contribution of torsion in this expression because we set it to zero at the end. However, one should keep those terms in the variation to consistently obtain the field equations. See \cite{Olmo:2013lta} for details.}  $\delta R_{\mu\nu}(\Gamma)=\nabla_{\lambda} \delta \Gamma_{\nu\mu}^{\lambda} - \nabla_{\nu} \Gamma_{\lambda \mu}^{\lambda}$ (valid for torsionless connections \cite{Olmo:2012vd}), and consider the piece of the variation in $\Gamma_{\mu\nu}^{\lambda}$ as
\be
\delta_{\Gamma} S= \frac{1}{2\kappa^2} \int d^d x \sqrt{-g} M_{\mu\nu}( \nabla_{\lambda} \delta \Gamma_{\nu\mu}^{\lambda} - \nabla_{\nu} \delta \Gamma_{\lambda \mu}^{\lambda})
\en
where we have introduced the object
\be\label{eq:M}
M^{\mu\nu}=f_R g^{\mu\nu}+2f_Q g^{\alpha\mu}g^{\beta\nu}R_{\alpha\beta}(\Gamma)
\en
Integrating by parts we obtain
\bea
\delta_{\Gamma} S &=& \frac{1}{2\kappa^2} \int d^d x \sqrt{-g} \Big[-\nabla_{\lambda} (\sqrt{-g} M^{\mu\nu}) \nonumber \\
&+& \frac{1}{2}\delta_{\lambda}^{\nu}\nabla_{\rho}(\sqrt{-g} M^{\mu\rho})+\frac{1}{2} \delta_{\lambda}^{\mu} \nabla_{\rho}(\sqrt{-g} M^{\nu\rho}) \Big] \delta \Gamma^{\lambda}_{\nu\mu}
\ena
Contracting $\lambda$ and $\nu$ in the expression above leads to $(1-d) \nabla_{\lambda} (\sqrt{-g} M^{\mu\lambda})=0$. In summary, the system of metric and connection equations for the action (\ref{eq:action}) in the Palatini formalism with the assumptions above becomes
\bea
f_R R_{\mu\nu}+2f_Q g^{\alpha\beta}R_{\mu\alpha}R_{\beta\nu}-\frac{1}{2}g_{\mu\nu}f&=&\kappa^2 T_{\mu\nu} \label{eq:metric} \\
\nabla_\lambda (\sqrt{-g} M^{\mu\nu})&=&0, \label{eq:connection}
\ena
where $T_{\mu\nu}=-\frac{2}{\sqrt{-g}} \frac{\delta S_m}{\delta g_{\mu\nu}}$ is the energy-momentum tensor of the matter.

To solve these equations we must first show that (\ref{eq:connection}) is linear in the connection. At first sight this is not obvious and, in fact, the definition (\ref{eq:M}) suggests that  (\ref{eq:connection}) is a nonlinear equation with up to second-order derivatives of the connection. The point is that one can use (\ref{eq:metric}) to show that the object ${P_\mu}^{\nu} \equiv R_{\mu\alpha}(\Gamma)g^{\alpha \nu}$ is an algebraic function of ${T_\mu}^\nu$, which allows to get rid of the connection dependence of $M^{\mu\nu}$ in favor of a dependence on the matter fields and the metric. To see this, we rise one index in (\ref{eq:metric}) with $g^{\mu\nu}$ to obtain
\begin{equation}\label{eq:metric1}
f_R {P_\mu}^{\nu}+2f_Q {P_\mu}^{\alpha}{P_\alpha}^{\nu}-\frac{f}{2}{\delta_\mu}^\nu=\kappa^2 {T_\mu}^\nu  \ .
\end{equation}
By noticing that $R= {P_\mu}^{\mu}$ and $Q={P_\mu}^{\nu}{P_\nu}^{\mu}$, it follows that ${P_\mu}^{\nu}$ must be an algebraic function of ${T_\mu}^{\nu}$. Eq.(\ref{eq:connection}) can thus be written as
\begin{equation}\label{eq:connection1}
\nabla_\lambda (\sqrt{-g} g^{\mu\alpha}{\Sigma_\alpha}^\nu)=0 \ ,
\end{equation}
where
\be\label{eq:Sigma}
{\Sigma_\alpha}^\nu\equiv f_R {\delta_\alpha}^{\nu}+2f_Q {P_\alpha}^{\nu}
\en
can be regarded as a function of the matter fields, i.e., $\hat P=\hat P(\hat T)$, with a hat denoting matrix representation of ${P_\alpha}^{\nu}$ and ${T_\alpha}^{\nu}$. In this form, the connection in (\ref{eq:connection1}) appears linearly and can be solved by means of elementary algebraic manipulations, in much the same way as the equation $\nabla_\lambda (\sqrt{-g} g^{\mu\alpha})=0$ is solved in the Palatini version of GR (see chapter 21 of \cite{MTW} for details). This can be seen by just proposing the existence of a rank-two tensor $h^{\mu\nu}$ such that
\begin{equation}\label{eq:lcconnection}
\sqrt{-g} g^{\mu\alpha}{\Sigma_\alpha}^\nu=\sqrt{-h}h^{\mu\nu} \ ,
\end{equation}
which turns (\ref{eq:connection1}) into $\nabla_\lambda (\sqrt{-h} h^{\mu\nu})=0$. This shows that the independent connection $\Gamma^\alpha_{\mu\nu}$ can be written as the Levi-Civita connection of the auxiliary metric $h_{\mu\nu}$. Note that this formal solution is valid for arbitrary Lagrangian $f(R,Q)$.

The explicit relation between $h_{\mu\nu}$ and $g_{\mu\nu}$ can be obtained by noting that $h=g \vert \hat{\Sigma} \vert^{\frac{2}{d-2}}$, where $\vert \hat{\Sigma} \vert$ represents the determinant of the matrix $\hat{\Sigma}$. Replacing this result back in (\ref{eq:lcconnection}) one obtains
\be \label{eq:connrelations}
\hat{h}=\vert \hat{\Sigma} \vert^{\frac{1}{d-2}} \hat{\Sigma}^{-1} \hat{g} \hspace{0.1cm};\hspace{0.1cm} \hat{h}^{-1}=\frac{\hat{g}^{-1} \hat{\Sigma}}{\vert \hat{\Sigma} \vert^{\frac{1}{d-2}}}
\en
The metric field equations (\ref{eq:metric}) can be conveniently written in terms of $h_{\mu\nu}$ by noting that (\ref{eq:metric1}) can be written as
\be \label{eq:metric2}
{P_\mu}^\alpha {\Sigma_\alpha}^\nu=\frac{f}{2}{\delta_\mu}^\nu+\kappa^2 {T_\mu}^\nu \ .
\en
Using the relations (\ref{eq:connrelations}), one easily verifies that  ${P_\mu}^\alpha {\Sigma_\alpha}^\nu= R_{\mu\alpha} (\Gamma) h^{\alpha \nu} \vert \hat{\Sigma} \vert^{\frac{1}{d-2}} =\frac{f}{2} {\delta_\mu}^{\nu} + \kappa^2 {T_\mu}^{\nu}$. Given that $R_{\mu\alpha}(\Gamma) =R_{\mu\alpha} (h)$, we finally get
\be \label{eq:Rmunu}
{R_\mu}^{\nu}(h)=\frac{\kappa^2}{\vert \hat{\Sigma} \vert^{\frac{1}{d-2}}} \left(\frac{f}{2\kappa^2} {\delta_\mu}^{\nu}+ {T_\mu}^{\nu} \right) \ ,
\en
where ${R_\mu}^{\nu}(h)=R_{\mu\alpha}h^{\alpha\nu}$.

Several comments are in order. First we note that this formal representation of the field equations recovers GR as a particular case when $f(R,Q)=R$, because in that case $\hat{\Sigma}=\hat{I}$ and $g_{\mu\nu}=h_{\mu\nu}$. On the other hand, it puts forward that the metric $h_{\mu\nu}$ satisfies a system of second-order field equations with the matter sources on the right hand side (recall that $R$ and $Q$ are functions of ${T_\mu}^\nu$). Since the physical metric $g_{\mu\nu}$ is algebraically related to $h_{\mu\nu}$ via the matter-induced {\it deformation} ${\Sigma_\alpha}^\beta$, it follows that $g_{\mu\nu}$ is also governed by second-order equations. To explore the properties of vacuum solutions, it is useful to complete the square in (\ref{eq:metric1}) and put it (in matrix notation, for simplicity) as
\begin{equation}\label{eq:mertic2}
(\hat P+\frac{f_R}{4f_Q}\hat I)^2=\left(\frac{f}{4f_Q}+\frac{f_R^2}{16f_Q^2}\right)\hat I+\frac{\kappa^2}{2f_Q}\hat T \ .
\end{equation}
In vacuum, $\hat T=0$, it follows that $\hat P=\alpha(R,Q) \hat I$, where the explicit form of $\alpha(R,Q)$ is not relevant for the discussion. From this relation one finds two constraints, $R\equiv \Tr[\hat P]=4\alpha(R,Q)$ and $Q\equiv\Tr[\hat P^2]=4\alpha(R,Q)^2$, which imply that  in vacuum $R$ and $Q$ must be constants. We thus find that ${\Sigma_\mu}^\nu\propto   {\delta_\mu}^{\nu}$, with the proportionality factor being a constant, which allows to write (\ref{eq:Rmunu}) as $R_{\mu\nu}(g)=\Lambda_{eff} g_{\mu\nu}$. This shows that regardless of the $f(R,Q)$ Lagrangian, the vacuum equations of the theory exactly recover the GR equations with an effective cosmological constant $\Lambda_{eff}$, whose form depends on the particular $f(R,Q)$ function chosen.

We have just seen that in the Palatini framework $f(R,Q)$ theories provide second-order modified dynamics with the same number of propagating degrees of freedom in vacuum as GR.  The presence of matter fields induces a nontrivial deformation between the metrics $g_{\mu\nu}$ and $h_{\mu\nu}$ which can have remarkable consequences in black hole scenarios, such as the replacement of the central singularity by a wormhole \cite{blor}. The second-order character of the field equations is guaranteed for arbitrary $f(R,Q)$ Lagrangian, polynomial or not, which contrasts with the tight constraints that appear in the usual metric formalism and motivate Lovelock theories and the quasi-topological models mentioned in the introduction. One is thus free to introduce new curvature couplings in the gravitational action without worries about the generation of ghosts or higher-order derivatives. The price to pay, however, is the need to solve algebraic equations required to express $R$ and $Q$ as functions of the matter fields and the coupling parameters. Some analytically tractable models of the $f(R)$ type have already been discussed in \cite{blor}. The introduction of a $Q-$dependence makes the analysis a bit harder, though some tractable {\it ad hoc} models can be found. In the next section we consider a better motivated theory which admits a complete analytical treatment and possesses all the algebraic properties found in the $f(R,Q)$ models presented here.

\section{Born-Infeld gravity in $d$ dimensions} \label{sec:III}

An interesting proposal for a high-energy gravitational action that recovers GR at low curvatures is that of Born-Infeld gravity. The idea is to construct a gravitational analog of the non-linear (classical) extension of Maxwell electrodynamics introduced by Born and Infeld \cite{BI}, where the Maxwell action is replaced by a square-root form. In this way, both the divergence of the field strength tensor and the self-energy associated to charged particles are removed within a classical model of the electromagnetic field. Deser and Gibbons \cite{Deser} considered its gravitational counterpart by essentially replacing the field strength tensor by the symmetric Ricci tensor obtaining
\bea \label{eq:actionBI}
S&=&\frac{1}{\kappa^2 \epsilon} \int d^d x \left[\sqrt{- \vert g_{\mu\nu}+ \epsilon R_{\mu\nu}(\Gamma) \vert } -\lambda \sqrt{-g} \right] \nonumber \\
&+& S_m(g_{\mu\nu},\psi_m)
\ena
where the same notation as in the previous section applies. The constant $\epsilon$ is a characteristic parameter of the theory with dimensions of length squared, and whose sign and value depend both on theoretical grounds and on its compatibility with experiments. This theory has attracted much interest in the last few years, with  applications in astrophysics \cite{BIa}, cosmology \cite{BIc}, and black hole scenarios \cite{ors}. This theory has also motivated new families of high-energy modifications of GR \cite{BIh}.

The meaning of the parameter $\lambda$ in (\ref{eq:actionBI})  follows from an expansion of the gravity sector, $S_{BI}$, in series of $\epsilon \to 0$ as
\bea \label{eq:quadratic}
\lim_{\epsilon\to 0} S_{BI}&=&\frac{1}{2\kappa^2} \int d^d x \sqrt{-g} [R-2\Lambda_{eff}] \\
&-&\frac{1}{2\kappa^2}\int d^d x \sqrt{-g} \frac{\epsilon}{2} \left(-\frac{R^2}{2} + R_{\mu\nu}R^{\mu\nu}\right) + \ldots \nonumber
\ena
where $\Lambda_{eff}=\frac{\lambda - 1}{\kappa^2 \epsilon}$ plays the role of the effective constant of the theory. This series expansion allows to see that BI gravity recovers the action of GR plus a cosmological constant in $d$ dimensions at low curvatures, with the next-to-leading order corresponding to that of a specific quadratic $f(R,Q)$ theory. Higher orders in this expansion involve contractions and products of increasing powers of the Ricci tensor.

The Lagrangian density appearing in (\ref{eq:actionBI}) can be conveniently rewritten by defining $q_{\mu\nu}\equiv g_{\mu\alpha}\left({\delta^\alpha}_\nu+\epsilon{P^\alpha}_\nu\right)$, where ${P^\alpha}_\nu\equiv g^{\alpha\beta}R_{\beta\nu}(\Gamma)$, and denoting ${\Omega^\alpha}_\nu\equiv g^{\alpha\beta}q_{\beta\nu}={\delta^\alpha}_\nu+\epsilon{P^\alpha}_\nu$. With this definitions, we have  $\hat\Omega={\hat g^{-1}}\hat q$ and $\hat\Omega^{-1}={\hat q^{-1}}\hat g$ (where a hat denotes matrix representation), which turns  (\ref{eq:actionBI}) into
\be \label{eq:BIr}
S_{BI}=\frac{1}{\kappa^2\epsilon}\int d^dx \sqrt{-g}\left[\sqrt{|\hat\Omega|}-\lambda \right]+S_m \ .
\en
This also suggests that BI gravity is a particular example of a more general family of theories, defined by the Lagrangian density $L_G=f(\vert \hat{\Omega} \vert)$, with $f=\sqrt{\vert \hat{\Omega} \vert}$ the BI case (see Ref.~\cite{or-extensions}).

Following the same procedure as in the $f(R,Q)$ case, variation with respect to metric and connection leads to the following field equations
\bea
\frac{\sqrt{-q}}{\sqrt{-g}} q^{\mu\nu}-\lambda g^{\mu\nu}&=&-\kappa^2 \epsilon T^{\mu\nu} \label{eq:metricBI} \\
\nabla_{\lambda}(\sqrt{-q} q^{\mu\nu})&=&0 \label{eq:connBI}  \ .
\ena
From the connection equation one readily sees that the connection is compatible with the rank-two tensor $q^{\mu\nu}$. Lowering an index in Eq.(\ref{eq:metricBI}) with $g_{\mu\nu}$ and using the relation $\hat\Omega^{-1}={\hat q^{-1}}\hat g$ introduced above,  it follows that (\ref{eq:metricBI}) can be rewritten as
\be \label{eq:Omega}
\vert \hat{\Omega} \vert^{1/2} {(\Omega^{-1})^\mu}_{\nu}=\lambda {\delta^\mu}_\nu -\epsilon \kappa^2 {T^\mu}_\nu
\en
This equation puts forward that, like in the $f(R,Q)$ case, the metric $g_{\mu\nu}$ and the metric $q_{\mu\nu}$ associated to the connection are related by a matrix $\hat \Omega$ which is fully determined by the matter sources. It is worth noting that defining $\hat{\Upsilon} \equiv \vert \hat{\Omega} \vert^{1/2} \hat \Omega^{-1}$, the relation between $g_{\mu\nu}$ and $q_{\mu\nu}$ take the form
\be \label{eq:gq}
q_{\mu\nu}=   \vert \hat{\Upsilon}\vert^{\frac{1}{d-2}} {(\Upsilon^{-1})_\mu}^{\alpha} g_{\alpha\nu} \hspace{0.1cm};\hspace{0.1cm}
q^{\mu\nu}=  \frac{1}{ \vert \hat{\Upsilon}\vert^{\frac{1}{d-2}}}  g^{\mu \alpha} {\Upsilon_\alpha}^{\nu} \ ,
\en
which are formally identical to those given in Eq.(\ref{eq:connrelations}) for the $f(R,Q)$ theories with $\hat{\Upsilon}$ playing the role of $\hat{\Sigma}$.

The field equations for the metric $q_{\mu\nu}$ can also be written in compact form. Starting with the definition of  $q_{\mu\nu}$ written in the form $\epsilon R_{\mu\nu}(\Gamma)=q_{\mu\nu}-g_{\mu\nu}$, we can raise an index with the metric $q^{\alpha \nu}$ to obtain $\epsilon {R_\mu}^{\nu}(q)={\delta_\mu}^{\nu} - {(\hat{\Omega}^{-1})_\mu}^{\nu}$ where ${R_\mu}^{\nu}(q) \equiv R_{\mu\alpha}(\Gamma)q^{\alpha \nu}$ is the Ricci tensor of the metric $q^{\mu\nu}$. Using now Eq.~(\ref{eq:Omega}), and the fact that $|\hat\Omega|^{1/2}=|\hat\Upsilon|^{\frac{1}{d-2}}$, we finally get
\be \label{eq:RmunuBI}
{R_\mu}^{\nu}(q)=\frac{\kappa^2}{|\hat\Upsilon|^{\frac{1}{d-2}}} \left[L_{BI} {\delta_\mu}^{\nu}+ {T_\mu}^{\nu} \right],
\en
where we have used the fact that the Lagrangian in the action (\ref{eq:actionBI}) can be written as
\be \label{eq:LBI}
L_{BI}=\frac{|\hat\Upsilon|^{\frac{1}{d-2}}-\lambda}{\kappa^2 \epsilon} \ .
\en
Remarkably, the form of the field equations (\ref{eq:RmunuBI}) for the BI theory, which constitutes a well motivated extension of GR, is identical with the form found for $f(R,Q)$ theories in (\ref{eq:Rmunu}), with the gravity Lagrangian in the $f(R,Q)$ case given by $L_{f}=f/2\kappa^2$.
This formal representation of the field equations is quite general in Palatini theories, though some slightly different examples are known \cite{Jimenez:2014fla,Makarenko:2014lxa}. Let us point out that, in the four dimensional case it has been observed that BI gravity coupled to an electromagnetic field coincides with the action of the quadratic theory, not only at the perturbative level (see Eq.~(\ref{eq:quadratic})), but to all orders, because the higher-order terms cancel out exactly due to the structure of the matter energy-momentum tensor \cite{ors}.

Like in the $f(R,Q)$ case,  Eq.(\ref{eq:RmunuBI}) constitutes a system of second-order field equations for the metric $q_{\mu\nu}$. Given that it is algebraically related via Eq.(\ref{eq:gq}) with $g_{\mu\nu}$, it follows that $g_{\mu\nu}$ also satisfies second-order equations.  It also confirms, that this framework naturally accommodates a cosmological constant term, as long as $\lambda \neq 1$. In vacuum, ${T_\mu}^{\nu}=0$, a glance at (\ref{eq:Omega}) confirms that the theory boils down to GR plus a cosmological constant term, in agreement with this general property of Palatini $f(R,Q)$ gravities.

\section{Electrovacuum solutions} \label{sec:IV}

In this section we shall work out black hole solutions of Born-Infeld gravity coupled to a spherically symmetric electric field in arbitrary dimensions, for which exact analytical solutions may be found.

\subsection{Action and field equations}

The matter (Maxwell) action in $d$ dimensions can be written as
\be \label{eq:matter}
S=-\frac{1}{16\pi} \int d^d x \sqrt{-g} F_{\mu\nu}F^{\mu\nu},
\en
where $F_{\mu\nu}=\partial_{\mu}A_{\nu}-\partial_{\nu}A_{\mu}$ is the field strength tensor of the vector potential $A_{\mu}$.
Let us assume a static space-time with a maximally symmetric $(d-2)$-dimensional subspace (which could be spherical, flat, or hyperbolic) with a line element of the form
\be \label{eq:line}
ds^2=g_{tt}dt^2+g_{xx}dx^2+r(x)^2 d\Omega^2_{(d-2)} \ ,
\en
where $d\Omega^2_{(d-2)}$%=d\theta_1^2+\sum_{i=2}^{n-1}\prod_{j=1}^{i-2} \sin^2 \theta_{j} d\theta_{i}^2$
denotes the infinitesimal line element of the maximally symmetric subspace and $r(x)^2, g_{tt}$, and $g_{xx}$ are functions of the spatial coordinate $x$. In the spherical case, $d\Omega^2_{(d-2)}=d\theta_1^2+\sum_{i=2}^{d-3}\prod_{j=1}^{i-2} \sin^2 \theta_{j} d\theta_{i}^2$. For static, spherically symmetric, electrovacuum solutions there is a single non-vanishing component, $F^{tx}(x)$, of the field strength tensor. From the Maxwell field equations, $\nabla_{\mu}F^{\mu\nu}=0$, this component in the line element (\ref{eq:line}) satisfies
\be\label{eq:Ftr}
F^{tx}=\frac{q}{r(x)^{d-2} \sqrt{-g_{tt}g_{xx}}}
\en
The energy-momentum tensor for an electromagnetic field, ${T_\mu}^{\nu}=-\frac{1}{4\pi} ({F_{\mu}}^{\alpha}{F_{\alpha}}^{\nu}-\frac{1}{4}{\delta_\mu}^{\nu} {F_\alpha}^{\beta}{F_{\beta}}^\alpha)$, defined by (\ref{eq:Ftr}) takes the form (from now on $n\equiv d-2$)
\be \label{eq:Tmunu}
{T_\mu}^{\nu}=\frac{X}{8\pi}
\left(
\begin{array}{cc}
-\hat{I}_{2\times 2}&  \hat{0}_{n \times 2} \\
\hat{0}_{2 \times n} & \hat{I}_{n \times n} \\
\end{array}
\right),
\en
where $X \equiv -\frac{1}{2}F_{\mu\nu}F^{\mu\nu}=\frac{q^2}{r(x)^{2(d-2)}}$. This ${T_\mu}^{\nu}$ does not depend explicitly on the metric components $g_{tt}$ and $g_{xx}$, which simplifies the resolution of the equations.

To fully specify the field equations in this scenario we need the explicit expression of $\hat{\Omega}$ following from Eq.(\ref{eq:Omega}), which in our case reads
\be \label{eq:Omegaemm}
\vert \hat{\Omega} \vert^{1/2} ({\hat{\Omega}^{-1})^\mu}_{\nu}=\left(
\begin{array}{cc}
(\lambda + \tilde{X}) \hat{I}_{2\times2} &  \hat{0}_{n \times 2} \\
\hat{0}_{2 \times n} & (\lambda - \tilde{X}) \hat{I}_{n \times n} \\
\end{array}
\right),
\en
where we have introduced the simplifying notation $\tilde{X} \equiv \frac{\epsilon \kappa^2}{8\pi} X$. Given the symmetry of our problem, we now propose the following ansatz
\be \label{eq:Omegaem}
\hat{\Omega}=
\left(
\begin{array}{cc}
\Omega_{+} \hat{I}_{2\times2} &  \hat{0}_{n \times 2} \\
\hat{0}_{2 \times n} & \Omega_{-} \hat{I}_{n \times n} \\
\end{array}
\right),
\en
Since the determinant of $\hat{\Omega}$ is $\vert \hat{\Omega} \vert = \Omega_{+}^2 \Omega_{-}^{n}$ we solve (\ref{eq:Omegaemm}) to obtain

\be \label{eq:omega}
\Omega_{-}=(\lambda + \tilde{X})^{\frac{2}{d-2}}  \hspace{0.1cm};\hspace{0.1cm} \Omega_{+}=\frac{(\lambda - \tilde{X} )}{(\lambda + \tilde{X})^{\frac{d-4}{d-2}}}
\en
The functions $\Omega_\pm$ determine the relative deformation between $g_{\mu\nu}$ and $q_{\mu\nu}$ and play a major role in the properties of the corresponding solutions. We now replace Eqs.(\ref{eq:Omegaem}) into the Lagrangian density (\ref{eq:LBI}), which becomes
\be \label{eq:LBI}
L_{BI}=\frac{\Omega_{+}\Omega_{-}^{\frac{d-2}{2}}-\lambda}{\epsilon \kappa^2 }.
\en
From (\ref{eq:LBI}),  (\ref{eq:Omegaem}) and (\ref{eq:Tmunu}) the field equations follow immediately after a bit of algebra:
\be \label{eq:Rmunuex}
\epsilon {R_\mu}^{\nu}(q)=
\left(
\begin{array}{cc}
\left(\frac{\Omega_{+}-1}{\Omega_{+}} \right) \hat{I}_{2\times2} &  \hat{0}\\
\hat{0} &  \left(\frac{\Omega_{-}-1}{\Omega_{-}} \right) \hat{I}_{n \times n} \\
\end{array}
\right),
\en
These field equations hold irrespective of the number of space-time dimensions and arise as an elegant extension of the four-dimensional counterparts studied in \cite{ors}.

\subsection{Formal solution of the field equations} \label{sec:s}

To solve the field equations (\ref{eq:Rmunuex}) we introduce a line element for the metric $q_{\mu\nu}$ of the form
\be \label{eq:metricq}
d\tilde{s}^2=-A(x)e^{2\psi(x)} dt^2 + \frac{1}{A(x)}dx^2 + \tilde{r}^2(x) d\Omega^2_n,
\en
where $\psi(x)$, $A(x)$, and $\tilde{r}(x)$ are three functions to be determined by the field equations. Note that $\tilde{r}(x)$ is, in general, a function of the radial coordinate $x$. With the line element (\ref{eq:metricq}) we can compute the components of the Ricci tensor in $d$ dimensions. The corresponding expressions have been worked out in detail in Appendix \ref{sec:app}. Equating such components to the right-hand-side of the field equations (\ref{eq:Rmunuex}) we find that ${R_t}^t-{R_x}^x=0$ which, from Eqs.(\ref{eq:Rtt}) and (\ref{eq:Rxx}), implies that $\tilde{r}_{xx}=\psi_x r_x$, which upon integration gives $\tilde{r}_x=e^{\psi}$. This allows us to introduce a new function $A\to {A}/\tilde{r}_x^2$ and redefine the $x$ coordinate as $\tilde{r}_x^2dx^2\to dx^2$ so that the line element may be written in a standard Schwarzschild form
\be
d\tilde{s}^2=-A(x)dt^2 +\frac{1}{A(x)} dx^2 + x^2 d\Omega_n^2 \ ,
\en
where now the role of $x$ as a radial coordinate is apparent. This leaves a single independent equation for $A(x)$, which from (\ref{eq:Rmm}) and (\ref{eq:Rmunuex}) reads
\be \label{eq:Rmn3}
\frac{\epsilon}{x^2} \left[(d-3)(k-A) -xA_x\right]= \left(\frac{\Omega_{-}-1}{\Omega_{-}} \right) \ ,
\en
where $k=-1,0,+1$ is the curvature of the maximally symmetric subspace. In this work we shall focus on the case $k=+1$ for simplicity (see however \cite{topological} for some examples of topological black holes).
We now take a mass ansatz in $d$ dimensions:
\be \label{eq:A}
A=1-\frac{2M(x)}{(d-3)x^{d-3}} \ ,
\en
where $M(x)$ is the mass function. Replacing this expression into equation (\ref{eq:Rmn3}) we find
\be \label{eq:Mass}
\frac{2\epsilon M_x}{(d-3)}=x^{d-2}\left(\frac{\Omega_{-}-1}{\Omega_{-}} \right).
\en
On the other hand, from the relation $\hat{\Omega}=\hat{g}^{-1}\hat{q}$ it follows that
\be \label{eq:qg}
q_{ab}=g_{ab} \Omega_{+} \hspace{0.1cm};\hspace{0.1cm} q_{mn}=g_{mn}\Omega_{-},
\en
where $(a,b)$ contains the $2\times2$ block and $(m,n)$ the maximally symmetric sector (see Appendix \ref{sec:app} for more details on the notation). It thus follows that the relation between the radial coordinate $x$ and the function $r^2(x)$ of the metric $g_{\mu\nu}$ is given by
\be \label{eq:coorrel}
x^2= r^2 \Omega_{-}.
\en
Taking a derivative in this equation, using the relation $\Omega_{-}^{\frac{d}{2}-1}=(\lambda + \tilde{X})$ and gathering terms, we obtain
\be \label{eq:coorel}
\frac{dx}{dr}=\Omega_{-}^{1/2} \left(\frac{\lambda - \tilde{X}}{\lambda + \tilde{X}} \right).
\en
Combining these results with Eq.(\ref{eq:Mass}) we get
\be \label{eq:Mr}
\frac{2\epsilon M_r}{(d-3)}=r^{d-2} \left(\frac{\Omega_{-}-1}{\Omega_{-}^{1/2}} \right) (\lambda -\tilde{X}).
\en
It is convenient at this point to introduce some simplifying notation. First, since we are mainly interested in the case $\epsilon<0$ (to meet the four-dimensional results studied in \cite{ors}) we redefine $\epsilon\equiv-l_{\epsilon}^2$ where $l_{\epsilon}$ is some length scale. The radial variable is rescaled as $z\equiv r/r_c$, where $r_c^{2(d-2)} \equiv l_{\epsilon}^2 r_q^{2(d-3)}$, with $r_q^{2(d-3)}\equiv\kappa^2 q^2/(4\pi)$ representing a length scale associated to the electric charge. With this notation, we get $\tilde{X}=-1/z^{2(d-2)}$, which is clearly dimensionless. With these elements we can integrate (\ref{eq:Mr}) as $M(z)=M_0 (1+ \delta_1 G(z)$), where $M_0$ is a constant identified with the Schwarzschild {\it mass} in vacuum, %(but with dimensions of length$^{d-3}$),
 $\delta_1$ is a dimensionless constant defined as
\be \label{eq:d1}
\delta_1 \equiv \frac{(d-3)r_c^{d-1}}{2 M_0 l_{\epsilon}^2} \ ,
\en
and the function $G(z)$, which accounts for the electric field contribution, satisfies
\be \label{eq:Gz}
G_z \equiv \frac{M_z}{\delta_1 M_0}=-z^{d-2} \left( \frac{\Omega_{-}-1}{\Omega_{-}^{1/2}} \right) \left( \lambda + \frac{1}{z^{2(d-2)}} \right) \ .
\en
Putting all these elements into (\ref{eq:A})  one arrives at the following formal expression
\be \label{eq:A(z)}
A(z)=1-\left(\frac{1+\delta_1 G(z)}{\delta_2  \Omega_{-}^{\frac{d-3}{2}}  z^{d-3}  }\right),
\en
where $\delta_2$ is another dimensionless constant given by
\be \label{eq:d2}
\delta_2 \equiv \frac{ (d-3)r_c^{d-3}}{2M_0}.
\en
To obtain the physical line element associated to $g_{\mu\nu}$, we use Eq.(\ref{eq:qg}), which yields
\be \label{eq:metricsol}
ds^2=-\frac{A}{\Omega_{+}} dt^2 +\left( \frac{\Omega_{+}}{A} \right) \left(\frac{dx}{\Omega_{+}} \right)^2 +z^2(x) d\Omega^2.
\en
The set of equations (\ref{eq:metricsol}), (\ref{eq:A(z)} and (\ref{eq:Gz}), with the definitions (\ref{eq:d1}) and (\ref{eq:d2}) and the area function $r^2(x)$ defined implicitly in (\ref{eq:coorrel}), provide a full solution for the problem of Born-Infeld gravity coupled to a spherically symmetric  electric field in arbitrary space-time dimensions.

\subsection{Exact solution and series expansions}

\begin{figure}[tbp]
\centering
\includegraphics[width=0.5\textwidth]{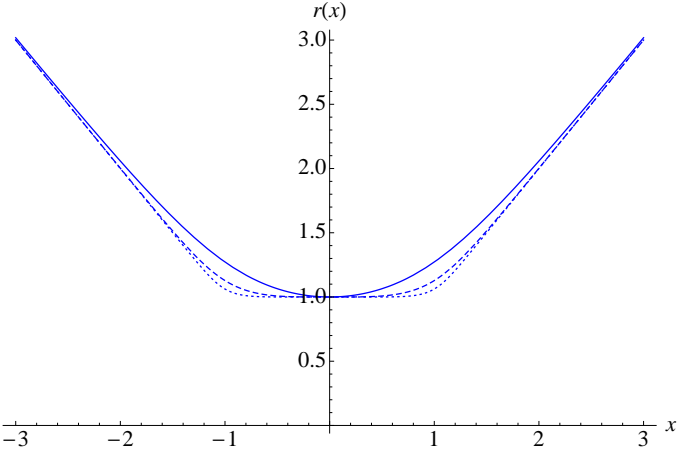}
\caption{Representation of $r(x)$ for $d=4$ (solid curve), $d=6$ (dashed curve), and $d=10$ (dotted curve).  The axes are measured in units of $r_c$.  The wormhole throat is located at $x=0$, where the area of the $(d-2)$-spheres reaches a minimum before bouncing into another space-time region. \label{fig:1}}
\end{figure}

In order to better understand some important aspects of these solutions, before considering the explicit resolution of (\ref{eq:Gz}) we show how to obtain an explicit expression for $r^2(x)$ in (\ref{eq:coorrel}). For simplicity we consider an asymptotically flat configuration, without cosmological constant, such that $\lambda=1$. The $\lambda\neq 1$ case has been considered in four dimensions \cite{Guendelman:2013sca}. Using the notation introduced in the above discussion, we can write (\ref{eq:coorrel}) as
\begin{equation}\label{eq:xz}
\frac{x^2}{r^2_c}=z^2\left(1-\frac{1}{z^{2(d-1)}}\right)^{\frac{2}{d-2}} \ .
\end{equation}
It is clear from this relation that $z\ge 1$, i.e., it can not become smaller than 1. The equality $z=1$ occurs when $x=0$. This has consequences for Eq.(\ref{eq:Gz}), which has a divergence at $z=1$. This is just a manifestation of the fact, well known in four dimensional models,  that the area function $z(x)\equiv r(x)/r_c$ is not a good coordinate at $x=0$. In fact, from (\ref{eq:coorel}) one can readily verify that $dr/dx$ vanishes at $x=0$ (or $r=r_c$), which signals a minimum in the function $r(x)$. As is well known, the area function $r(x)$ can only  be used as a coordinate in those intervals in which it is a monotonic function \cite{Stephani2003}. To cover that point, therefore, one should use the coordinate $x$, for which the function $G(x)$ is well behaved. Now, with simple manipulations,  (\ref{eq:xz}) can be written as
\begin{equation}\label{eq:xz1}
\left(\frac{|x|}{r_c}\right)^{(d-2)}=\frac{1}{z^{d-2}}\left(z^{2(d-2)}-1\right) \ ,
\end{equation}
which is just a standard quadratic algebraic equation for $z^{d-2}$. The solution is straightforward and leads to (see Fig.\ref{fig:1})
\begin{equation}\label{eq:r(x)}
r^{d-2}=\frac{|x|^{d-2}+\sqrt{|x|^{2(d-2)}+4r_c^{2(d-2)}}}{2}
\end{equation}
The modulus in the coordinate $x$ follows from the fact that it appears as $x^2$ in the original relation (\ref{eq:coorrel}) and because the range of definition of $x$ can be extended to the whole real axis. This gives consistency to our spherically symmetric, sourceless electric field, whose effective charge now appears as a topological property of the electric flux through a wormhole (see also Sec.\ref{sec:flux} below). The throat of this wormhole is located at $x=0$, where the area function $r(x)$ reaches its minimum value.

Now that the domains of the radial coordinate $x$ and the area function $r(x)$ have been clarified, we proceed to solve (\ref{eq:Gz}) to obtain $G(z)$. We will focus on the asymptotically flat case $\lambda=1$. Using the binomial expansion of the different functions that appear on the right-hand side of this equation, an exact solution in terms of infinite power series can be obtained for arbitrary dimension. In general, the solution takes the form ($d\ge 4$)
 \begin{eqnarray}
G_d(z)&=& -\frac{z^{(3-d)}}{(d-3)} {_2F_1}\left[\frac{1}{d-2},\frac{d-3}{2d-4},\frac{3d-7}{2d-4};\frac{1}{z^{2d-4}}\right] \nonumber \\
&+&\frac{z^{(d-1)}}{d-1}{_2F_1}\left[\frac{1}{d-2},-\frac{(d-1)}{2d-4},\frac{d-3}{2d-4};\frac{1}{z^{2d-4}}\right]\nonumber \\&-&\frac{z^{(d-1)}}{d-1}(1-z^{4-2d})^{(d-1)} \ ,
\end{eqnarray}
where ${_2F_1}$ is a hypergeometric function. One can verify that in the limit $z\to \infty$ the GR result is recovered, $G(z)\approx -\frac{2}{(d^2-5d+6)}\frac{1}{ z^{d-3}}$. The other limit of interest is when $z\to 1$. In that case, we find
\begin{equation}
\lim_{z\to 1} G_d(z)\approx -\frac{1}{\delta_d}+a_d(z-1)^{\frac{d-3}{d-2}}+b_d(z-1)^{\frac{d-1}{d-2}}
\end{equation}
where $\delta_d$, $a_d$ and $b_d$ are constants given by
\begin{eqnarray}\label{eq:delta_d}
\delta_d^{-1}&=&\frac{-\pi\csc\left(\frac{\pi}{d-2}\right)}{(d-1)\Gamma\left[\frac{1}{d-2}\right]}\left(\frac{\Gamma\left[\frac{d-3}{2d-4}\right]}{\Gamma\left[\frac{d-5}{2d-4}\right]}-\frac{(d-1)}{(d-3)}\frac{\Gamma\left[\frac{3d-7}{2d-4}\right]}{\Gamma\left[\frac{3(d-3)}{2d-4}\right]}\right) \\
a_d&=& \frac{\pi\csc\left[\frac{\pi}{d-2}\right]2^{\frac{d-3}{d-2}}(d-2)^{-\frac{1}{d-2}}}{\Gamma\left[\frac{1}{d-2}\right]\Gamma\left[\frac{d-5}{d-2}\right]}\\
 b_d&=& -\frac{(2(d-2))^{\frac{d-1}{d-2}}}{d-1} \ .
\end{eqnarray}
where $\Gamma[a]$ is Euler's gamma function. The expansion of the metric function (\ref{eq:A(z)}) near the wormhole throat becomes
\begin{eqnarray}
A(z)&\approx & \frac{(\delta_1-\delta_d)}{\delta_2\delta_d}\frac{(2(d-2))^{-\frac{d-3}{d-2}}}{(z-1)^{\frac{d-3}{d-2}}}\left(1+\frac{(d-3)}{(2(d-2)}(z-1)\right) \nonumber\\
&+&1-\frac{\delta_1(2(d-2))^{-\frac{d-3}{d-2}}}{\delta_2} \left(a_d-\frac{2b_d(d-2)}{(d-1)}(z-1)^{\frac{2}{d-2}}\right.\nonumber\\
&+&\left.\frac{2a_d(d-3)}{(2d-5)}(z-1)+\frac{2b_d}{(d-1)}(z-1)^{\frac{d}{d-2}}\right) \ .
\end{eqnarray}
One can readily verify that this function diverges at $z=1$ if $\delta_1\neq \delta_d$ but is completely smooth if $\delta_1=\delta_d$. For the regular case, the  corrections that follow the zeroth-order constant term depend on the dimension, being linear in $(z-1)$ in $d=4$ but going like $(z-1)^{\frac{2}{d-2}}$ in higher dimensions. Similarly as in $d=4$, depending on the ratio $\delta_1/\delta_2$ one might find configurations without event horizons which, roughly, can be interpreted as two ($d-$dimensional) Minkowski space-times connected through a wormhole.

We can now consider the form of the metric component $g_{tt}=-A(z)/\Omega_+$ around the region $z=1$, which takes the form
\begin{eqnarray}\label{eq:gttz}
g_{tt}&\approx & -\frac{(2(d-2))^{\frac{d-4}{d-2}}}{2}\left[\frac{(\delta_1-\delta_d)}{\delta_2\delta_d}\frac{(2(d-2))^{-\frac{d-3}{d-2}}}{(z-1)^{\frac{1}{d-2}}} \right. \nonumber\\
&+& \left(1-\frac{a_d\delta_1(2(d-2))^{-\frac{d-3}{d-2}}}{\delta_2} \right)(z-1)^{\frac{d-4}{d-2}} \nonumber\\
&+& \frac{(\delta_1-\delta_d)}{\delta_2\delta_d}\frac{(4d-7)}{(2(d-2))^{\frac{2d-5}{d-2}}}(z-1)^{\frac{d-3}{d-2}}\nonumber\\
&+& \left.\frac{\delta_1b_d}{\delta_2} \frac{(2(d-2))^{\frac{1}{d-2}}}{(d-1)}(z-1)\right]
\end{eqnarray}
For $\delta_1\neq \delta_d$, the first term in this expression provides the main contribution. In terms of the coordinate $x$, near $x\to 0$ it turns into
\begin{equation} \label{eq:gttx}
g_{tt}\approx -\frac{(\delta_1-\delta_d)}{2\delta_2\delta_d}\frac{r_c}{|x|} \ .
\end{equation}
For $\delta_1= \delta_d$, we find instead
\begin{eqnarray}
g_{tt}&\approx&-\frac{1}{2}\left(1-\frac{a_d\delta_1(2d-4)^{-\frac{d-3}{d-2}}}{\delta_2} \right)\left(\frac{|x|}{r_c}\right)^{d-4}\nonumber\\&-&\frac{b_d\delta_d}{\delta_2(d-1)(2(d-2))^{\frac{1}{d-2}}}\left(\frac{|x|}{r_c}\right)^{d-2} \ .\label{eq:gttxreg}
\end{eqnarray}
From this last expansion, it is easy to see that only in $d=4$ can the near-wormhole geometry be free of curvature divergences, as any derivative of the metric near $x\to 0$ will introduce inverse powers of $|x|$ even in the regular case. Unlike in $d=4$, no horizonless solutions can exist for $d>4$ even if $\frac{a_d\delta_1(2d-4)^{-\frac{d-3}{d-2}}}{\delta_2}<1$. In that case, however, the event horizon is not hiding any interior region, as it only exists at the wormhole throat, acting in this way as a kind of membrane that separates two asymptotically Minkowskian space-times.

\subsection{Curvature scalars}\label{sec:Krets}

To better understand the properties of curvature in these geometries, we now compute the Kretschmann scalar,  $K=R_{\alpha\beta\gamma\delta}R^{\alpha\beta\gamma\delta}$ (see Appendix \ref{sec:app2}) around the region $x=0$ where the deviations with respect to GR occur.  For arbitrary values of the parameters $\delta_1, \delta_2$ we can use Eq.(\ref{eq:gttx}) together with (\ref{eq:KretD}) and (\ref{eq:Kret2}) to expand in series around $x=0$ ($z=1$). We first check the far limit behavior, $z \gg 1$, which yields $K_d \sim 1/r^{4(d-2)} $, in agreement with the behaviour of the Kretschmann scalar in the RN case of GR. On the other hand, for $x \simeq 0$, the corresponding result admits a generic expression
\be \label{eq:KD}
K_d \simeq \frac{A_d(\delta_1-\delta_d)^2}{\delta_2^2 \delta_d^2 x^{4d-10}} + \frac{B_d(\delta_1,\delta_2,\delta_d)}{x^{3d-7}} + \ldots
\en
where $A_d, B_d,\ldots$ are constants multiplying each term in the series. The expansion around $x \simeq 0$ can be easily transformed into an expansion in $z$ just by using (\ref{eq:r(x)}), which behaves there as $x\simeq (z-1)^{1/(d-2)}$. %(note that in the standard RN case of GR one finds $K_{RN} \simeq 1/r^{4(d-2)}$.
The expansion in (\ref{eq:KD}) shows that the degree of divergence grows with $d$. In the leading-order term we always find a factor $(\delta_1-\delta_d)$, which means that the degree of divergence is softened in the case $\delta_1=\delta_d$. Replacing this choice in the metric expressions, and performing again series expansions around $x \simeq 0$ yields
\be
K_d|_{\delta_1=\delta_d} \simeq \frac{C_d(\delta_2,\delta_d)}{x^{2(d-2)}} + \frac{E_d(\delta_2,\delta_d)}{x^{2(d-3)}} + \ldots
\en
where $C_d, E_d, \ldots$ is a set of new constants. This expansion is still divergent for $d>4$ (in terms of $z$ one finds the same degree of divergence, namely, $\simeq C_d/(z-1)^2$), though it can be further smoothed out if the ratio $\delta_2/\delta_d$ is properly chosen, but divergences still persist. However, for $d=4$, the $\delta_1=\delta_d$ case leads to finiteness of the Kretschmann (and other curvature scalars), since one obtains $C_4=E_4=0$ and
\be
K_4|_{\delta_1=\delta_d}  \simeq \left(16-\frac{64\delta_1}{3\delta_2} + \frac{88 \delta_1^2}{9\delta_2^2} \right) + O(x^2) \ ,
\en
which is a result already found in the literature \cite{or}.

\subsection{Electric flux and action integral}\label{sec:flux}

As pointed out above, from the analysis of the radial function $r(x)$ given by Eq.(\ref{eq:r(x)}) (see Fig.\ref{fig:1}) we concluded that our geometry represents a wormhole space-time. We can reinforce this interpretation by exploring in some detail the origin of the electric charge associated to our solutions. The electric charge can be defined via the flux integral
\be
\Phi=\int_{S^{d-2}} *F = \omega_{d-2} q \ ,
\en
where  $*F$ is the Hodge dual to the electromagnetic field tensor, and the integration takes place over a boundary defined by a $(d-2)$ closed hypersphere $S^{d-2}$, with $\omega_{d-2} = 2\pi^{\frac{d-1}{2}}/\Gamma[\frac{d-1}{2}]$ representing the volume of a unit $(d-2)$ sphere. The hypersurface $S^{d-2}$  contains the minimal sphere $x=0$ in its interior and, for this reason, the net flux is non-zero and yields a finite charge $q$. The hypersphere can be chosen to be on the side $x>0$ or on the side $x<0$. If its orientation is chosen in the direction of growth of the $(d-2)-$spheres, then the charge estimated by an observer in $x>0$ has different sign from that computed by an observer in $x<0$. This type of solutions were first described by Wheeler and Misner as the result of the combination of free electric fields with non-trivial topologies \cite{Wheeler}, giving rise to the notion of {\it mass without mass and charge without charge}.

An important property of these solutions is that the area of the wormhole, $A=\omega_{D-2} r_c^{d-2}$, grows with the intensity of the electric flux in such a way that the electric field at the throat is a universal constant independent of the particular charge or mass characterizing the solution. In fact, dividing the electric flux by the area of the wormhole throat, we find
\be
\frac{\Phi}{\omega_{d-2} r_c^{d-2}}= \frac{\sqrt{\kappa^2/4\pi}}{l_\epsilon} \ ,
\en
where, recall, $l_{\epsilon}$ is the scale characterizing the Born-Infeld Lagrangian and $\kappa^2$ is proportional to Newton's constant in $d$ dimensions.  This puts forward the existence of topological properties (associated with the wormhole structure) which are independent of the particular geometrical aspects of the solutions. Note that in the limit $l_\epsilon\to 0$, where GR is recovered and the wormhole shrinks to zero area, the above quantity diverges.

Another interesting aspect to note is the fact that the spatial integral of the action describing these solutions is finite. Without knowing the explicit form of the solutions, it is possible to evaluate the integral in (\ref{eq:BIr}) to obtain a generic result that signals another universal property of these solutions. This integral can be written explicitly as
\begin{eqnarray}
S_T&=&\frac{1}{\kappa^2\epsilon}\int dt dx\frac{ r^{(d-2)} d\omega_n }{\Omega_+}\left[|\hat\Omega|^{1/2}-1+\frac{\kappa^2\epsilon X}{8\pi}\right] \\
&=& \frac{2\omega_{d-2} r_c^{d-1}}{\kappa^2\epsilon}\int dt \int_1^\infty \frac{dz z^{d-2}(1-\tilde{X})}{(1+\tilde{X})^{\frac{1}{(d-2)}}}\left[(1+\tilde{X})^{\frac{2}{(d-2)}}-1\right] \nonumber
\end{eqnarray}
Defining $\kappa^2\equiv \frac{2\omega_{d-2} G_d}{(d-3)c^3}$, where $G_d$ is the $d-$dimensional Newton's constant and $c$ the speed of light, restoring factors of $G_d$ and $c$ in (\ref{eq:A}) to make it dimensionally consistent (with $M_0$ a mass), and using the definition of $\delta_1$, we can rewrite the factor in front of the integrals as $ \frac{2\omega_{d-2} r_c^{d-1}}{\kappa^2\epsilon}=-2M_0 c^2 \delta_1$.  Evaluation of the integral  gives the highly nontrivial result
\begin{equation}
S_T=2M_0c^2\frac{\delta_1}{\delta_d}\int dt \ ,
\end{equation}
where $\delta_d$ is the coefficient defined in (\ref{eq:delta_d}) resulting from the series expansion of the function $G(z)$ in the limit $z\to 1$ (or $x\to 0$). This result is a rather generic property of this type of theories and is not restricted to the coupling of gravity with Maxwell's electrodynamics \cite{Olmo:2013mla}. The wormhole structure has been essential to regularize the radial integral, which diverges in GR because in that case the integration extends down to $0$ in its lower range. The factor $2$ multiplying the mass is due to the need to integrate on both sides of the wormhole. Interestingly, the resulting action coincides with that of a relativistic point particle of mass $2M_0\frac{\delta_1}{\delta_d}$ at rest.

\section{Geodesics} \label{sec:V}

We have seen above in Sec.\ref{sec:Krets} that curvature divergences arise generically at the wormhole throat, which are frequently regarded as a sign of the existence of space-time singularities. However, singular space-times are actually characterized by the existence of inextendible paths \cite{Geroch:1968ut,Curiel2009}. In particular, if there exist null or time-like geodesics which cannot be extended to arbitrarily large values of their affine parameters, i.e., if they have a beginning or an end, then the space-time can be regarded as singular. In this section we address this important point of the geometries presented above. A comprehensive discussion of the case $d=4$ appears in \cite{GCiWHST}.

Taking advantage of the high degree of symmetry of our spherical space-times, instead of solving the geodesic equation, we work directly with the norm of the tangent vector of geodesics and its conserved quantities. Denoting the geodesic paths as $\gamma^\mu(\tau)=(t(\tau),x(\tau),\theta_i(\tau))$, conservation of angular momentum implies that we have $(d-2)$ relations of the form $L_i=g_{ii} \frac{\text{d}\theta_i}{\text{d}\tau}$. Because of spherical symmetry, we can rotate the coordinates so that the first $(d-3)$ angles are fixed at $\pi / 2$ and all the motion takes place in the last angle, $\theta_{d-2}\equiv \varphi$, so the conserved quantity that matters is the total angular momentum $L=r^2\frac{\text{d}\varphi}{\text{d}\tau}$. Time translation invariance gives another conserved quantity, namely, $E=-g_{tt}\frac{\text{d}t}{\text{d}\tau}$. The radial component of the geodesic tangent vector is thus governed by
\begin{equation}\label{eq:geodx}
 \frac{1}{\Omega_+}\frac{\text{d}x}{\text{d}\tau} = \pm\sqrt{E^2+g_{tt}\left ( \eta + \frac{L^2}{r^2(x)} \right )}  \ ,
\end{equation}
where $\eta=0,1$ for null or time-like geodesics, respectively. The $\pm$ sign in (\ref{eq:geodx}) denotes outgoing/ingoing geodesics.  From this last equation, we see that the motion in the radial direction is analog to that of a particle of energy $\mathcal{E}=E^2$ in a one dimensional potential

\begin{equation} \label{eq:Veff}
V_{eff}=-g_{tt}\left ( \eta + \frac{L^2}{r^2(x)} \right ).
\end{equation}

\begin{figure}[tbp]
\centering
\includegraphics[width=0.5\textwidth]{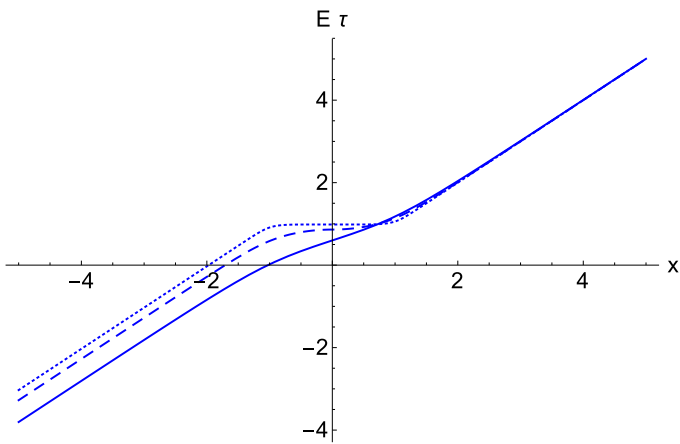}
\caption{Representation of the affine parameter $E \tau(x)$ as a function of the radial coordinate $x$ for outgoing radial null geodesics in $d=4$ (solid curve), $d=5$ (dashed curve), and $d=10$ (dotted curve). The smooth curves show that radial null geodesics can be naturally extended through the wormhole throat, $x=0$, where curvature scalars generically diverge. \label{fig:NullRad}}
\end{figure}

For null radial geodesics, $\eta=0=L^2$, the geodesic equation admits an exact analytical expression in the form
\begin{equation}
\pm E \tau(x)=\left\{\begin{array}{lr} W[\lambda, d; z] & \text{ if } x>0 \\  & \\
2C[\lambda,d] -W[\lambda, d; z] & \text{ if } x<0 \end{array}\right. \ ,
\end{equation}
where
\begin{eqnarray}
W[\lambda, d; z]&=&
\frac{r_c}{\lambda }z \left(\lambda -z^{4-2 d}\right)^{\frac{d-3}{d-2}} \times \\ & &_2F_1\left(1,1+\frac{3}{4-2 d},1+\frac{1}{4-2 d};\frac{z^{4-2 d}}{\lambda }\right) \nonumber \ ,
\end{eqnarray}
and $_2F_1(a,b,c; x)$ is a hypergeometric function. In the case $\lambda=1$, the constant $C[\lambda,d]=W[\lambda,d;1]$ takes the form
\begin{equation}
C[1,d]=\frac{\pi  \csc \left(\frac{\pi  (d-3)}{d-2}\right) \Gamma \left(\frac{1}{4-2 d}\right)}{3 \Gamma \left(\frac{3}{4-2 d}\right) \Gamma \left(\frac{1}{d-2}\right)} \ .
\end{equation}
A plot of $E\tau(x)$ is shown in Fig.\ref{fig:NullRad}, from which one readily sees that the affine parameter can be smoothly extended across the wormhole throat regardless of the parameters characterizing the solution.

We now consider time-like and null geodesics with $L^2\neq 0$ and focus on those which can get very near the wormhole throat to explore their extendibility through the region with curvature divergences. We will thus use the approximations given in (\ref{eq:gttx}) and (\ref{eq:gttxreg}) for $g_{tt}$ in $V_{eff}$ (see Eq.(\ref{eq:Veff})). These paths can be classified in three groups, depending on the value of $\delta_1$:
\begin{itemize}
 \item If $\delta_1>\delta_d$, the potential near the wormhole is an infinite barrier, $V_{eff}\sim 1/|x|$, and the geodesics do not reach the wormhole. In these configurations, the wormhole is in a time-like region (below the inner horizon) and, therefore, these geodesics can be extended to arbitrary values of their affine parameter  in much the same way as in GR.

 \item If $\delta_1<\delta_d$, the potential is an infinite well, and the geodesic reaches the wormhole. Near the wormhole $g_{tt} \propto \frac{1}{|x|}$ [see (\ref{eq:gttx})], and $\Omega_+ \approx 2 |x|^{(4-d)}$. It is possible to integrate (\ref{eq:geodx}) in the neighbourhood of the wormhole to obtain $\pm \sqrt{\frac{2r_c(\delta_d-\delta_1)}{\delta_2\delta_d}\left(\eta+\frac{L^2}{r_c^2}\right)}(\tau-\tau_0)\approx \frac{x|x|^{(d-7/2)}}{d-5/2}$. Since $x\in]-\infty,+\infty[$, it is clear that the range of the proper time $\tau$ extends over the whole real line.
 \item If $\delta_1=\delta_d$, $g_{tt}$ is either proportional to $|x|^{(d-4)}$ or to $|x|^{(d-2)}$ as the wormhole is approached and $V_{eff}$ can be neglected if $d>4$ (see \cite{GCiWHST} for a discussion of the $d=4$ case). We thus find that  $\pm E(\tau-\tau_0)\approx \frac{x|x|^{(d-4)}}{2(d-3)}$, which also allows the affine parameter to smoothly go through the $x=0$ region.
\end{itemize}

We have thus seen that in all those cases in which geodesics can reach the wormhole, the affine parameter $\tau(x)$ can always be extended through it to cover its whole domain of definition over the real line. We also note that the integration of $\frac{d\varphi}{d\tau}=L/r^2(x)$ leads to a well-defined and continuous $\varphi(\tau)$ because $r(x)\to r_c>0$ as $x=0$, being the integrand finite everywhere. Integration of the coordinate $t(\tau)$ is also smooth and gives the result $(t-t_c)\propto (\tau-\tau_0)|\tau-\tau_0|^{1/(d-5/2)}$ if $\delta_1<\delta_d$, or $(t-t_c) \propto \tau-\tau_0$ if $\delta_1=\delta_d$, where $t_c$ is the value of the coordinate $t$ at the point of crossing the wormhole.
These results confirm that the existence of curvature divergences at $x=0$ does not prevent in any way the extendibility of geodesic paths, which puts forward that these space-times are nonsingular.

\section{Conclusions} \label{sec:VI}

In this work we have considered metric-affine (or Palatini) extensions of GR including powers of the curvature scalar, $R$, and the Ricci tensor in an arbitrary number of space-time dimensions. This scenario generalizes that of five-dimensional $f(R)$ theories studied in \cite{blor}, and that of four-dimensional $f(R)$ \cite{or11}, $f(R,Q)$ theories \cite{or}, and Born-Infeld gravities \cite{ors}. We have explicitly shown that the field equations are always second-order and ghost-free, as they naturally recover the GR equations in vacuum. This property is a result of the independence of the connection with respect to the metric. These theories, therefore, provide an appealing new scenario to investigate aspects of the AdS/CFT correspondence.

We have shown that the connection satisfies a set of algebraic, non-dynamical equations that result from the variation of the action with respect to it. The solution to those equations implies that an auxiliary metric $q_{\mu\nu}$ can be defined such that the independent connection coincides with the Christoffel symbols of $q_{\mu\nu}$. The field equations for $q_{\mu\nu}$ can then be cast in Einstein-like form, with the right-hand side of the equations representing a kind of modified matter source, whose form depends on the particular gravity Lagrangian chosen. As the physical metric $g_{\mu\nu}$ is algebraically related to $q_{\mu\nu}$ via the matter sources, finding a solution for $q_{\mu\nu}$ automatically provides a solution for $g_{\mu\nu}$.

We have illustrated our analysis employing the gravitational analogue of Born-Infeld theory of non-linear electrodynamics, dubbed Born-Infeld gravity. It has been shown that the field equations for this theory admit a similar formal representation as those of $f(R,Q)$ theories, which confirms the robustness of this formulation. This theory naturally accommodates  a non-vanishing cosmological constant as long as $\lambda \neq 1$. We have obtained analytical solutions for Born-Infeld gravity coupled to a spherically symmetric electrovacuum field, generalizing in this way the Reissner-Nordstr\"{o}m solution of GR. Remarkably, we have found that the central point-like singularity of GR is generically replaced by a wormhole structure. Performing series expansions around the wormhole throat, where the modified gravitational dynamics becomes important, we have shown that all solutions with $d>4$ exhibit curvature divergences at the throat. These divergences, however, do not prevent null and time-like geodesics from being extendible across the wormhole, which has a highly nontrivial implication, namely,  that the resulting space-times are nonsingular. Interestingly, though the wormhole size grows with the intensity of the electric flux (number of charges), it does in such a way that the electric field at the throat is a universal constant. On the other hand, the wormhole structure of these solutions plays a key role in the regularization of the action, whose spatial integral is finite.

The results presented in this work, therefore, put forward that black hole singularities in arbitrary dimensions can be cured in a purely classical geometric scenario governed by second-order field equations. The use of metric-affine extensions of Einstein's theory has been fundamental to get this type of modified (but second-order) dynamics. The fact that space-times with curvature divergences can be nonsingular is in sharp contrast with the widely spread strategy of bounded scalars to obtain nonsingular space-times [see e.g. \cite{Mukhanov:1991zn,Ansoldi:2008jw,Lemos,Spallucci,Bronnikov,Hayward} and references therein], and demands for a reconsideration of the role typically attributed in the literature to curvature divergences to characterize space-time singularities.  In our case, the emergence of wormholes connecting two copies of the space-time, which give structure to the standard GR point-like singularities, has been crucial to guarantee geodesic completeness in all cases. We point out that the presence of such wormholes seems to be a generic prediction of metric-affine geometries, since they arise in four-dimensional $f(R)$ and $f(R,Q)$ theories as well \cite{OR-U,or,ors}. Further analysis has revealed that the wormhole structure in such space-times implies, besides geodesic completeness, preservation of causal contact for  physical observers going through the wormhole\cite{Congruences}, and a well behaved (scalar) wave propagation \cite{Olmo:2015dba} across the divergent region. Such an analysis should be extended to the scenario considered in this work. The robustness of these results in similar scenarios with other matter sources will be explored elsewhere. Extensions of these results to models involving  further curvature invariants, such as powers of the Riemann tensor with couplings different from those appearing in the Gauss-Bonnet/Lovelock theories, will require the development of new methods to deal with the connection equations. We hope to address all these issues in future works.

\section*{Acknowledgments}

D.B. and L.L. would like to thank CAPES and CNPq for financial support. G.J.O. is supported by a Ramon y Cajal contract, the Spanish grant FIS2011-29813-C02-02, the Consolider Program CPANPHY-1205388,  and  the i-LINK0780 grant of the Spanish Research Council (CSIC). D.R.-G. is supported by the NSFC (Chinese agency) grants No. 11305038 and 11450110403, the Shanghai Municipal Education Commission grant for Innovative Programs No. 14ZZ001, the Thousand Young Talents Program, and Fudan University. The authors also acknowledge funding support of CNPq project No. 301137/2014-5.

\appendix

\section{Decomposition of the Ricci tensor} \label{sec:app}

In this appendix we consider the expressions for the Ricci tensor corresponding to a metric of the form

\be \label{eq:gmunu}
g_{\mu\nu}^{(D)}=
\left(
\begin{array}{cc}
g_{ab}^{(2\times 2)} &  \hat{0} \\
0 & g_{mn}^{(n \times n)}  \\
\end{array}
\right),
\en
where the spherical sector satisfies $g_{mn}^{n\times n}=r^2(x^a)\eta_{mn}(\theta^i)$. Here the following conventions apply: $a, b, c, d, e$ are indices corresponding to the $2 \times 2$ sector, while $i, j, l, m, n$ are indices for the spherical sector. $D$ is the total number of space-time dimensions and $n=D-2$ the spherical sector dimensionality.  The coefficients of the (Levi-Civita) connection for this metric read

\bea \label{eq:apcon}
\Gamma^{bam}&=&0 \hspace{0.1cm};\hspace{0.1cm} \Gamma^{b}_{mn}=-g^{al}r\partial_l r \eta_{mn} \hspace{0.1cm};\hspace{0.1cm} \nonumber \\
\Gamma^i_{cd}&=&0\hspace{0.1cm};\hspace{0.1cm} \Gamma^{i}_{jc}=\frac{\partial_c r}{r} \delta_j^i
\ena

From the definition of the Riemann tensor (\ref{eq:Riemann}) we obtain the Ricci tensor
%(note that no metric structure is needed to do this)

\be
R_{\beta \nu} \equiv R^{\alpha}_{\beta \alpha \nu}=\partial_{\alpha}\Gamma^{\alpha}_{\nu\beta}-\partial_{\nu}\Gamma^{\alpha}_{\alpha\beta}+ \Gamma^{\alpha}_{\alpha\lambda}\Gamma^{\lambda}_{\nu\beta} - \Gamma^{\alpha}_{\nu\lambda} \Gamma^{\lambda}_{\alpha\beta}
\en
and we decompose it into its $2\times 2$ and spherical sectors as follows

\bea
R_{cd}^{(D)} &=&R_{cd}^{(2)}-(D-2)\frac{\nabla_a \nabla_c r}{r}  \label{eq:Ricci2x2}\\
R_{mn}^{(D)}&=&R_{mn}^{(2)}-\eta_{mn} [r \Box r + (D-3)g^{ac} \nabla_a r \nabla_r ] \label{eq:Riccinxn}
\ena
For the $2 \times 2$ sector  we consider a static, spherically symmetric metric, which can be cast as

\be \label{eq:q2t2}
g_{ab}^{(2\times 2)}=
\left(
\begin{array}{cc}
-Ae^{2\psi} &  0\\
0&  \frac{1}{A} \\
\end{array}
\right),
\en
From (\ref{eq:Ricci2x2}) we obtain for this case

\bea
R_{tt}^{(2)}&=&\frac{A^2 e^{2\psi}}{2} \left[\frac{3A_x \psi_x}{A} + \frac{A_{xx}}{A} + 2(\psi_{xx}+\psi_x^2) \right] \\
R_{xx}^{(2)}&=&-\frac{1}{2} \left[\frac{3A_x \psi_x}{A} + \frac{A_{xx}}{A} + 2(\psi_{xx}+\psi_{x}^2) \right]
\ena
and therefore it follows that ${R_t}^{(2)t}={R_x}^{(2)x}$.

On the other hand, concerning the $(D-2)$ sector, for a maximally symmetric space we have $R^{\alpha}_{\beta\mu\nu}=k(\delta_{\mu}^{\alpha}\eta_{\beta \nu} - \delta_{\nu}^{\alpha} \eta_{\beta \mu})$ (with $k=1,0,-1$), which implies $R_{\beta\nu}^{(D-2)}=(D-3)k\eta_{\beta\nu}^{(D-2)}$. Thus, from (\ref{eq:Riccinxn}) we have

\be
R_{mn}^{(D)}=\eta_{mn}^{(D-2)} \left[(D-3)(k-g^{ac}\nabla_a r \nabla_c r)-\Box r\right]
\en
and raising an index gives $R_m^{n(D)}=\frac{\delta_m^n}{r^2}[(D-3)(k-g^{ac}\nabla_a r \nabla_c r )-\Box r]$. To make these expressions explicit we note that for the metric (\ref{eq:q2t2}) we have the components of the connection $\Gamma_{tt}^x=\frac{A^2 e^{2\psi}}{2}(\frac{A_x}{A}+2\psi_x)$ and $\Gamma_{xx}^x=-\frac{A_x}{2A}$ we find that

\bea
\nabla_t\nabla_t r&=&-\frac{A^2 e^{2\psi}}{2}r_x \Big(\frac{A_x}{A}+2\psi_x  \Big) \\
\nabla_x \nabla_x r&=&r_{xx}+\frac{A_x}{2A}r_x \\
\Box r &=&A\Big[r_{xx}+\frac{A_x}{A}r_x+\psi_x r_x \Big]
\ena
Putting all these elements together we arrive to the equations

\bea
{R_t}^t&=& -\frac{A}{2} \Big[ \frac{A_{xx}}{A}+3\frac{A_x}{A} \psi_x+2(\psi_x^2 + \psi_{xx}) \nonumber \\
&+&(D-2)\Big(\frac{r_x}{r} \frac{A_x}{A} + 2\psi_r \frac{r_x}{r} \Big) \Big] \label{eq:Rtt}\\
{R_x}^x&=&  -\frac{A}{2} \Big[ \frac{A_{xx}}{A}+3\frac{A_x}{A} \psi_x+2(\psi_x^2 + \psi_{xx}) \nonumber \\
&+&(D-2)\Big(\frac{r_x}{r} \frac{A_x}{A} + 2 \frac{r_{xx}}{r} \Big) \Big] \label{eq:Rxx} \\
{R_m}^n&=& \frac{\delta_m^n}{r^2} \Big[(D-3) (k-Ar_x^2) \nonumber \\
&-&Ar \Big(r_{xx}+r_x \Big(\frac{A_x}{A}+\psi_x \Big) \Big) \Big] \label{eq:Rmm}
\ena
which are the ones employed in section (\ref{sec:s}).

\section{The Kretschmann scalar in $D$ dimensions} \label{sec:app2}

From the components of the connection (\ref{eq:apcon}), associated to a metric of the form (\ref{eq:gmunu}), the non-vanishing components of the Riemann tensor (\ref{eq:Riemann}) read
\bea
R^{a(D)}_{bcd}&=&R^{a(2)}_{bcd} \nonumber  \hspace{0.1cm};\hspace{0.1cm} R^{a(D)}_{ibj}=-\eta_{ij}rg^{ae}\nabla_b\nabla_e r =-R^{a(D)}_{ijb}\\
R^{i(D)}_{jmn}&=&R^{i(n)}_{jmn} +(\delta_n^i \eta_{mj}-\delta_m^i \eta_{nj} ) g^ {ae}\nabla_a r \nabla_e r \nonumber \\
R^{i(n)}_{jmn}&=&k(\delta_m^i \eta_{jn} - \delta_n^i \eta_{jm})  \\
R^{i(D)}_{ajb}&=&-\delta_j^i \frac{\nabla_a \nabla_b r}{r}=-R^{i(D)}_{abj} \nonumber %\\
%R^{a(D)}_{ijb}&=& \eta_{ij}g^{al} r \nabla_b \nabla_e r \hspace{0.1cm};\hspace{0.1cm} R^{i(D)}_{bcj}=\delta_j^i \frac{\nabla_c \nabla_b r}{r}
\nonumber
\ena
The Kretschmann scalar is written as
\bea
K^{(D)}&=&{R^{\alpha (D)}}_{\beta\mu\nu}{R_\alpha}^{\beta \mu\nu (D)} \nonumber \\
&=&({R^{a(D)}}_{bcd}{R_a}^{bcd(D)} + 2 {R^{a(D)}}_{ibj}{R_a}^{ibj(D)})  \\
&+&({R^{i(D)}}_{jkl}{R_i}^{jkl(D)} + 2{R^{i(D)}}_{bja}{R_i}^{bja(D)}) \nonumber
\ena
Computing each contribution separately one obtains
\bea
{R^{a(D)}}_{bcd}{R_a}^{bcd(D)}&=&K^{(2)} \\
{R^{a(D)}}_{ibj}{R_a}^{ibj(D)}&=&(D-2)\frac{\nabla_a\nabla_b r}{r}\frac{\nabla^a \nabla^b r}{r} \\
{R^{i(D)}}_{jmn}{R_i}^{jmn(D)}&=&{R^{i(n)}}_{jmn}{R_i}^{jmn(n)} \nonumber \\
&-&4kC(D-2)(D-3) \\
&+&2C^2(D-2)(D-3) \nonumber \\
{R^{i(n)}}_{jkl}{R_i}^{jkl(n)} &=& 2k^2 (D-2)(D-3) \\
{R^{i(D)}}_{bja}{R_i}^{bja(D)}&=&(D-2) \frac{\nabla_a\nabla_b r}{r}\frac{\nabla^a \nabla^b r}{r}
\ena
where $C=g^{ab}\nabla_a r \nabla_b r$. Therefore, the Kretschmann scalar reads
\bea \label{eq:Kret}
K^{(D)}&=&K^{(2)} + 4(D-2)\left(\frac{\nabla_a \nabla_b r}{r} \right)\left(\frac{\nabla^a \nabla^b r}{r} \right) \nonumber \\
&+& \frac{2(D-2)(D-3)(k-g^{ab}\nabla_a r \nabla_b r)^2}{r^4}
\ena
Consider now the $2 \times 2$ sector written as
\be \label{eq:gab}
g_{ab}^{(2)}=
\left(
\begin{array}{cc}
-A(x) &  0 \\
0 & B(x)  \\
\end{array}
\right),
\en
to meet the expression of the metric $g_{ab}$, corresponding to an electromagnetic field, considered in this work. For this metric the relevant objects appearing in (\ref{eq:Kret}) read
\be
(\nabla_a\nabla_b r)(\nabla^a \nabla^b r)=\frac{1}{B^2} \left[\frac{ A^2_x}{4A^2}r_x^2 + \left(r_{xx}-\frac{B_x}{2B}r_x \right)^2 \right]
\en
Then, the Kretschmann (\ref{eq:Kret}) reads explicitly:
\bea \label{eq:KretD}
K^{(D)}&=&K^{(2)} + \frac{4(D-2)}{r^2B^2} \left[\frac{ A^2_x}{4A^2}r_x^2 + \left(r_{xx}-\frac{B_x}{2B}r_x \right)^2 \right]  \nonumber\\
&+& \frac{2(D-2)(D-3)}{r^4} \Big(k-\frac{r_x^2}{B} \Big)^2
\ena
where
\be \label{eq:Kret2}
K^{(2)}=\frac{1}{4B^2} \left[\frac{A_x}{A} \frac{B_x}{B} + \left(\frac{A_x}{A}\right)^2 -2\frac{A_{xx}}{A} \right]^2
\en
For the particular case of the line element (\ref{eq:metricsol}) we can write $A(x)=C(x)/\Omega_{+}(x)$ and $B(x)=1/(C(x)\Omega_{+}(x))$. Taking into account the relations $\Omega_{+,x}=\Omega_{+,r} r_x$, $\Omega_{-,xx}=\Omega_{+,rr}r_x^2 + \Omega_{+,r}r_{xx}$, $ A_r=C_r/\Omega_{+} - C \Omega_{r}/\Omega_{+}^2$ and $B_r=-C_r/(C^2 \Omega_{+}) - \Omega_r^+/(C \Omega_{+}^2)$ one can compute the Kretschmann scalar (\ref{eq:KretD}), where
\be
K^{(2)}=\frac{1}{\Omega_{+}^2} \left[C\Omega_{+,x}^2 + \Omega_{+}C_{xx}-\Omega_{+} (C_x \Omega_{+,x}+C\Omega_{+,xx}) \right]^2
\en
and the full expression for $K^{(D)}$ is immediately obtained by replacing the above expressions into (\ref{eq:KretD}).


\begin{thebibliography}{99}

\bibitem{AdS/CFT}
J. Maldacena, Adv. Theor. Math. Phys. {\textbf{2}}, 231 (1998);
Int. J. Phys. {\textbf{38}}, 1113 (1999);
E. Witten, Adv. Theor. Math. Phys. {\textbf{2}}, 253 (1998);
S. S. Gubser, I. R. Klebanov, and A. M. Polyakov, Phys. Lett. B {\textbf{428}}, 105 (1999).

\bibitem{Condensed}
S. Sachdev, Lect. Notes Phys. \textbf{828}, 273 (2011).

\bibitem{Dictionaries}
S. R. Das, T. Nishioka, and T. Takayanagi, JHEP \textbf{1007}, 071 (2010);
D.~Harlow and D.~Stanford, \emph{Operator Dictionaries and Wave Functions in AdS/CFT and dS/CFT}, arXiv:1104.2621 [hep-th]

\bibitem{GB}
D. L. Wiltshire, Phys. Rev. D \textbf{38}, 2445 (1988);
M. Aiello, R. Ferraro, and G. Giribet, Phys. Rev. D \textbf{70}, 104014 (2004);
M. H. Dehghani, N. Alinejadi, and S. H. Hendi, Phys. Rev. D \textbf{77}, 104025 (2008);
M. Aiello, R. Ferraro, and G. Giribet, Class. Quant. Grav. \textbf{22}, 2579 (2005);
H. Maeda, M. Hassaine, and C. Martinez, Phys. Rev. D \textbf{79}, 044012 (2009);
S. H. Hendi, Phys. Lett. B \textbf{677}, 123 (2009);
D. Rubiera-Garcia, Phys. Rev. D \textbf{91}, 064065 (2015).

\bibitem{Lovelock}
D. Lovelock, J. Math. Phys. \textbf{12}, 498 (1971);
C. Charmousis, Lec. Notes Phys. \textbf{769}, 299 (2008);
C. Garraffo and G. Giribet, Mod. Phys. Lett. A \textbf{23}, 1801 (2008).

\bibitem{quasi-topological}
R. C. Myers and B. Robinson, JHEP {\textbf{1008}}, 067 (2010);
R. C. Myers, M. F. Paulos, and A. Sinha, JHEP {\textbf{1008}}, 035 (2010);
W. G. Brenna and R. B. Mann, Phys. Rev. D {\textbf{86}}, 064035 (2012);
M.~H.~Dehghani and M.~H.~Vahidinia, JHEP {\bf 1310}, 210 (2013) 210;
U. Camara da Silva, G.M. Sotkov, Nuclear Physics B \textbf{874}, 471 (2013);
R.~A.~Hennigar, W.~G.~Brenna and R.~B.~Mann, arXiv:1505.05517 [hep-th].

\bibitem{crystalline}
C. Kittel, Introduction to Solid State Physics (8th edition), Wiley, 2005;
E. Kr\"oner,
%\emph{The differential geometry of elementary point and line defects in Bravais crystals},
Int. J. Theor. Phys. \textbf{29}, 1219 (1990);
R. de Witt,
%\emph{A view of the relation between the continuum theory of lattice defects and non Euclidean geometry in the linear approaximation},
Int. J. Engng. Sci. \textbf{19}, 1475 (1981);
F. Falk,
%\emph{Theory of elasticity of coherent inclusions by means of nonmetric geometry},
J. Elast. \textbf{11}, 359 (1981).

\bibitem{orl14}
F. S. N. Lobo, G. J. Olmo, and D. Rubiera-Garcia, arXiv:1412.4499 [hep-th]

\bibitem{Zanelli}
J. Zanelli,
\emph{Lecture notes on Chern-Simons (super-) gravities},
arXiv:hep-th/0502193.

\bibitem{Exirifard}
Q. Exirifard and M. M. Sheikh-Jabbari, Phys. Lett. B \textbf{661}, 158 (2008).

\bibitem{Borunda}
M. Borunda, B. Janssen, and M. Bastero-Gil, JCAP \textbf{0811}, 008 (2008).

\bibitem{Olmo:2011uz}
  G.~J.~Olmo,
  %``Palatini Approach to Modified Gravity: f(R) Theories and Beyond,''
  Int.\ J.\ Mod.\ Phys.\ D {\bf 20}, 413 (2011).

\bibitem{blor}
D. Bazeia, L. Losano, G. J. Olmo, and D. Rubiera-Garcia, Phys. Rev. D \textbf{90}, 044011 (2014).



\bibitem{Curiel2009}
E. Curiel and P. Bokulich, {\it Singularities and Black Holes}, The Stanford Encyclopedia of Philosophy (\href{http://plato.stanford.edu/archives/fall2012/entries/spacetime-singularities/}{Fall 2012 Edition}), Edward N. Zalta (ed.).

\bibitem{Geroch:1968ut}
  R.~P.~Geroch,
  %``What is a singularity in general relativity?,''
  Ann. Phys.\  {\bf 48}, 526 (1968).

\bibitem{Hawking:1973uf}
  S. W. Hawking and G. F. R. Ellis, \emph{The Large Scale Structure of Space-Time} (Cambridge University Press,
Cambridge, 1973).

\bibitem{Wald:1984rg}
  R.~M.~Wald, \emph{General Relativity} (Chicago, Usa: University Press, 1984).


\bibitem{GCiWHST}
G.~J.~Olmo, D.~Rubiera-Garcia and A.~Sanchez-Puente, {\it Geodesic completeness in a wormhole space-time with horizons}, to appear (2015).

\bibitem{Congruences}
G.~J.~Olmo, D.~Rubiera-Garcia and A.~Sanchez-Puente, {\it Impact of curvature divergences on physical observers in a wormhole space-time with horizons}, to appear (2015).

\bibitem{Olmo:2015dba}
  G.~J.~Olmo, D.~Rubiera-Garcia and A.~Sanchez-Puente,
  %``Classical resolution of black hole singularities via wormholes,''
  arXiv:1504.07015 [hep-th].

\bibitem{Olmo:2013lta}
  G.~J.~Olmo and D.~Rubiera-Garcia,
  %``Importance of torsion and invariant volumes in Palatini theories of gravity,''
  Phys.\ Rev.\ D {\bf 88}, 084030 (2013).


\bibitem{Olmo:2012vd}
  G.~J.~Olmo,
  %``Cosmology in Palatini theories of gravity,''
  AIP Conf.\ Proc.\  {\bf 1458}, 222 (2011).

\bibitem{MTW}
C. W. Misner, K. S. Thorne, and J. A. Wheeler, \emph{Gravitation and cosmology} (W. H. Freeman, 1973).

\bibitem{BI}
M. Born and L. Infeld, Proc. R. Soc. London A \textbf{144}, 425 (1934).
\bibitem{Deser}
S. Deser and G. W. Gibbons, Class. Quant. Grav. {\bf 15}, L35 (1998);
M. Ba\~nados and P. G. Ferreira, Phys. Rev. Lett. {\bf 105}, 011101 (2010).

\bibitem{BIa}
P. Pani, V. Cardoso, and T. Delsate, Phys. Rev. Lett. \textbf{107}, 031101 (2011).
P.~Pani and T.~P.~Sotiriou, Phys.\ Rev.\ Lett.\  {\bf 109}, 251102 (2012).

\bibitem{BIc}
J. H. C. Scargill, M. Ba\~nados, and P. G. Ferreira, Phys. Rev. D \textbf{86}, 103533 (2012);
T. Harko, F. S. N. Lobo, M. K. Mak, and S. V. Sushkoc, Mod. Phys. Lett. A \textbf{29},  1450049 (2014).
M. Ba\~nados, Phys. Rev. D \textbf{77}, 123534 (2008);
M. Ba\~nados, P. G. Ferreira, and C. Skordis, Phys. Rev. D \textbf{79}, 063511 (2009);
P. P. Avelino and R. Z. Ferreira, Phys. Rev. D \textbf{86}, 041501 (2012).

\bibitem{ors}
G. J. Olmo, D. Rubiera-Garcia, and H. Sanchis-Alepuz, Eur. Phys. J. C \textbf{74}, 2804 (2014).

\bibitem{BIh}
P. Pani, T. Delsate, and V. Cardoso, Phys. Rev. D \textbf{85}, 084020 (2012).

\bibitem{or-extensions} S.D. Odintsov, G. J. Olmo, and D. Rubiera-Garcia, Phys. Rev. D \textbf{90}, 044003 (2014).


\bibitem{Jimenez:2014fla}
  J.~B.~Jiménez, L.~Heisenberg and G.~J.~Olmo,
  %``Infrared lessons for ultraviolet gravity: the case of massive gravity and Born-Infeld,''
  JCAP \textbf{1411}, 004 (2014).

\bibitem{Makarenko:2014lxa}
  A.~N.~Makarenko, S.~Odintsov and G.~J.~Olmo,
  %``Born-Infeld-$f(R)$ gravity,''
  Phys.\ Rev.\ D {\bf 90}, 024066 (2014).

  \bibitem{topological}
L.~Vanzo,
  %``Black holes with unusual topology,''
  Phys.\ Rev.\ D {\bf 56}, 6475 (1997);
D.~Klemm,
  %``Topological black holes in Weyl conformal gravity,''
  Class.\ Quant.\ Grav.\  {\bf 15}, 3195 (1998);
 R.~-G.~Cai, J.~-Y.~Ji and K.~-S.~Soh,
  %``Topological dilaton black holes,''
  Phys.\ Rev.\ D {\bf 57}, 6547 (1998);
A.~Sheykhi,
  %``Topological Born-Infeld-dilaton black holes,''
  Phys.\ Lett.\ B {\bf 662}, 7 (2008);
R.~B.~Mann,
  %``Lifshitz Topological Black Holes,''
  JHEP {\bf 0906}, 075 (2009).


\bibitem{Guendelman:2013sca}
  E.~I.~Guendelman, G.~J.~Olmo, D.~Rubiera-Garcia and M.~Vasihoun,
  %``Nonsingular electrovacuum solutions with dynamically generated cosmological constant,''
  Phys.\ Lett.\ B {\bf 726}, 870 (2013).

\bibitem{Stephani2003}
H. Stephani, D. Kramer, M. Maccallum, C. Hoenselaers, and E. Herlt, {\it Exact solutions of Einstein's field equations} (Cambridge University Press, 2003).

\bibitem{or}
G. J. Olmo and D. Rubiera-Garcia, Phys. Rev. D \textbf{86}, 044014 (2012);
Eur. Phys. J. C \textbf{72}, 2098 (2012);
Int. J. Mod. Phys. D \textbf{21}, 1250067 (2012).

\bibitem{Wheeler}
J. Wheeler, Phys. Rev. D \textbf{97}, 511 (1955);
C. W. Misner and J. A. Wheeler, Ann. Phys. \textbf{2}, 525 (1957).


\bibitem{Olmo:2013mla}
  G.~J.~Olmo and D.~Rubiera-Garcia,
  %``Semiclassical geons at particle accelerators,''
  JCAP {\bf 1402}, 010 (2014).


\bibitem{or11} G. J. Olmo and D. Rubiera-Garcia, Phys. Rev. D \textbf{84}, 124059 (2011).


\bibitem{Mukhanov:1991zn}
  V.~F.~Mukhanov and R.~H.~Brandenberger,
  %``A Nonsingular universe,''
  Phys.\ Rev.\ Lett.\  {\bf 68}, 1969 (1992).

\bibitem{Ansoldi:2008jw}
  S.~Ansoldi,
  %``Spherical black holes with regular center: A Review of existing models including a recent realization with Gaussian sources,''
  arXiv:0802.0330 [gr-qc];

\bibitem{Lemos}
J.~P.~S.~Lemos and V.~T.~Zanchin,
  %``Regular black holes: Electrically charged solutions, Reissner-Nordstr\'om outside a de Sitter core,''
  Phys.\ Rev.\ D {\bf 83}, 124005 (2011).

\bibitem{Spallucci}
E.~Spallucci, A.~Smailagic, and P.~Nicolini,
  %``Non-commutative geometry inspired higher-dimensional charged black holes,''
  Phys.\ Lett.\ B {\bf 670},  449 (2009).

\bibitem{Bronnikov}
K.~A.~Bronnikov and J.~C.~Fabris,
  %``Regular phantom black holes,''
  Phys.\ Rev.\ Lett.\  {\bf 96}, 251101 (2006).

\bibitem{Hayward}
S.~A.~Hayward,
  %``Formation and evaporation of regular black holes,''
  Phys.\ Rev.\ Lett.\  {\bf 96},  031103 (2006).

\bibitem{OR-U}
G. J. Olmo and D. Rubiera-García, {\it Nonsingular black holes in $f(R)$ theories}, to appear (2015).

\end{thebibliography}
\end{document}